\documentclass[12pt]{amsart}
\usepackage{amssymb,array}

\newtheorem{lemma}{Lemma}

\newtheorem{theorem}[lemma]{Theorem}
\newtheorem{corollary}[lemma]{Corollary}

\newcommand{\CC}{\mathcal{C}}
\newcommand{\CA}{\mathcal{A}}

\newcommand{\R}{\mathbb{R}}
\newcommand{\C}{\mathbb{C}}

\newcommand{\N}{\mathbb{N}}

\newcommand{\dom}{\mathrm{Dom}\,}

\newcommand{\ud}{\,\mathrm{d}}

\title[Heat kernel of the CSNW]{The heat kernel of the
compactified $D=11$ supermembrane with non-trivial winding}
\author[L.~Boulton, A.~Restuccia]
{L.~Boulton$^1$, A.~Restuccia$^2$}

\date{17$^\mathrm{th}$ February 2005}

\thanks{$^1$PIMS postdoctoral fellow.}

\keywords{Regularized hamiltonian, compactified $D=11$ supermembrane with
non-trivial winding, heat kernel, matrix Feynman-Kac formula.}

\begin{document}
\begin{abstract}
We study the quantization of the regularized hamiltonian, $H$, of
the compactified $D=11$ supermembrane with non-trivial winding. By
showing that $H$ is a relatively small perturbation of the bosonic
hamiltonian, we construct a Dyson series for the heat kernel of $H$
and prove its convergence in the topology of the von
Neumann-Schatten classes so that $e^{-Ht}$ is ensured to be of
finite trace. The results provided have a natural interpretation in
terms of the quantum mechanical model associated to regularizations
of compactified super\-membranes. In this direction, we discuss the
validity of the Feynman path integral description of the heat kernel
for $D=11$ supermembranes and obtain rigorously a matrix Feynman-Kac
formula.
\end{abstract}

\maketitle

\section{Introduction}

The $D=11$ supermembrane with non-trivial central charge (CSNW) was
first analyzed in the semiclassical regime in \cite{duff} and
\cite{od}. A study of the complete theory, including all the
interacting terms in the hamiltonian, was performed later in
\cite{1,5,2} where it was shown that the spectrum of the $SU(N)$
regularized hamiltonian is a discrete set of eigenvalues of finite
multiplicity. In order to have a non-trivial central charge when the
spatial part of the world volume is compact, the supermembrane must
wrap a compact sector of the target space. This is achieved by
imposing a topological restriction on the configuration space. These
restrictions are naturally satisfied in the context of brane
wrapping calibrated submanifolds, \cite{B1,B2,B3}. As it turns out,
the ground state configuration of the CSNW is directly related to
the intersecting brane solutions of $D=11$ supergravity,
\cite{A1,A2,A3,A4,S+T,A6}.

In the present paper we continue our analysis of the quantum
hamiltonian of the CSNW. The approach discussed below aims at a
better understanding of the rigorous behaviour of the  regularized
hamiltonian of $D=11$ supermembranes in the limit $N\to \infty$. An
important step towards the visualization of this limit is provided
by Theorem~\ref{t1} below.

Two are the properties of the CSNW that are crucial in allowing a
detailed description the spectrum of the $SU(N)$ regularized
hamiltonian, $H$, \cite{5,1}. One is the fact that $H$ is a small
perturbation of the bosonic hamiltonian $H_B=P^2+V_B$, cf.
\eqref{e8} below. The other is the certainty that the bosonic
potential $V_B$ is bounded below by $cQ^2$ for suitable $c>0$, see
Lemma~\ref{t4} below. Our main objective in the present paper is to
demonstrate how these properties also lead to an explicit
description on the heat kernel of $H$, see Theorem~\ref{t1} and
Corollary~\ref{t3}.

The main contribution is to be found in Section~4. The heat kernel
of $P^2+Q^2$, the harmonic oscillator, is given by Mehler's formula
\begin{equation} \label{e9}
   \mathrm{Ker}[e^{-(P^2+Q^2)t}](x,y)=
   (w_1/\pi)^{N/2}\exp[2w_1(x\cdot y)-w_2(|x|^2+|y|^2)]
\end{equation}
where
$w_1=\frac{\lambda}{1-\lambda^2}$, $w_2=\frac{1+\lambda^2}
 {2(1-\lambda^2)}$, and $\lambda=\exp[-2 t]$.
An explicit computation shows that $\mathrm{Ker}[e^{-(P^2+Q^2)t}]\in
L^2(\R^N\times \R^N)$. By dilating the spatial variable, it is easy
to show that the latter also hold for
$K^A_t:=\mathrm{Ker}[e^{-(P^2+cQ^2)t}]$. If we now compare
$K^B_t:=\mathrm{Ker}[e^{-(P^2+V_B)t}]$ with $K^A_t$ through the
Feynman-Kac formula, it becomes clear that also $K^B_t$ is square
integrable so that the c$_0$ one-parameter semigroup $e^{-H_Bt}$ is
Hilbert-Schmidt and hence (by the semigroup property) of finite
trace for all $t>0$. On the other hand, the results of \cite{1,2}
ensure that $H$ realizes as the Schr\"odinger-type operator
$P^2+V_B\otimes \mathbb{I}+V_F$, acting on $L^2(\R^N;\C^{d})$ for
suitably large dimensions $N,d\in \N$, where
$V_F:\R^N\longrightarrow \R^{d\times d}$ is subordinated to
$V_B\otimes \mathbb{I}$ in a sense specified below. Here the spinors
are vectors in $\C^d$ so that $\mathbb{I}$ denotes the identity
matrix of $\C^{d\times d}$. In Theorem~\ref{t1} below we demonstrate
the convergence for all $t>0$ of the Dyson expansion for $e^{-Ht}$
in the norm of the von Neumann-Schatten classes of index $r$ for a
suitable $r>2$. Therefore, also $e^{-Ht}$ has a finite trace.

The series expansion found for $e^{-Ht}$ has a natural physical
interpretation. If we formally consider the Feynman functional
integral for the CSNW and perform the gauge fixed $SU(N)$
regularization procedure described in \cite{2}, we achieve a Feynman
path integral with light cone coordinates denoted by $x^+$. Since
the full potential, $V_S:=V_B\otimes \mathbb{I}+V_F$, is bounded
below, this formula represents the kernel of the Schr\"odinger
operator of a quantum mechanical model with a light cone time whose
hamiltonian is $H$. The heat kernel is the analytic continuation
from $x^+$ onto $-ix^+$ of the above Feynman path integral. Thus, it
is of physical relevance constructing this heat kernel for finite
$N$ and considering then the limit as $N\to \infty$. This limit may
have an intrinsic relevance in the description of the CSNW.

Section~5.1 is devoted to describing $K_t:=\mathrm{Ker}[e^{-Ht}]$ as
a Feynman path integral in terms of the standard Wiener measure.
Although this description provides some information about the model,
in our view, a deeper insight into the quantization of the CSNW is
obtain from the abovementioned comparison between the fermionic and
the bosonic sectors of the theory. Following this approach, and
using the construction of Section~4, in Section~5.2 we obtain an
alternative Feynman-Kac formulation.  The latter is given in terms
of a measure in the space of continuous path constructed \textit{a
la} Wiener, but using the kernel of $e^{-H_Bt}$, instead of the free
heat kernel, for  defining the measure of cylinder sets. This
procedure follows the approach suggested in \cite[p.49]{glja} for
the scalar case. In conjunction with the analysis of Section~4,
this formula is highly relevant in the construction of a Feynman
functional integral formula in the limit $N\to \infty$.

It is interesting to notice that the results of the remarkable
papers \cite{6,deWit+H+N,deWit+M+N} are somehow in contrast with the
present discussion. The regularized hamiltonian of the supermembrane
immerse in $D=11$ Minkowski target space has spectrum equal to the
interval $[0,\infty)$, \cite{6}. If we do not compactify with
non-trivial wrapping certain directions in the target space, the
existence of locally singular configurations, also known as
string-like spikes, force the whole regularized hamiltonian to be
incompatible with the bosonic term. By virtue of the spectral
theorem, the evolution semigroup can not be a compact operator.
Furthermore, in this case the hamiltonian is positive, but since the
fermionic potential is not subordinated to the bosonic contribution,
a Feynman path integral construction of the heat kernel is still an
open problem.

In order to keep our discussion accessible as far as functional
analytical properties of $H$ is concerned, Section~3 is devoted to
reviewing some standard mathematical tools employed in the
subsequent parts of this paper. In Section~2 we briefly justify why
$H$ is the correct characterization of a quantum mechanical
hamiltonian for the CSNW model. We refer to \cite{1,2,3,4}  for a
more complete account that such a representation is valid. In
\cite{1,2} we quantize the CSNW by first solving the constraint and
the gauge fixing condition, and then performing a canonical
reduction of the hamiltonian. For the present discussion we quantize
the CSNW model by imposing the constrains on the Hilbert space of
states. This second procedure yields to the expression $H=P^2+V_S$
considered here. It is well known that both schemes of quantization
are equivalent.

\section{The regularized hamiltonian of the CSNW}

Supermembranes \cite{S+T} are extended objects which live in 11
dimensions and may couple to 11 dimensional supergravity. These
objects were originally proposed over a $D=11$ Minkowski target
space as candidates for fundamental objects. When the possibility of
non-empty continuous spectrum was discovered in \cite{6}, this lead
to a reinterpretation of the model as a many body theory. The
continuous nature of the spectrum is a consequence of two main
facts: the presence of singular configurations, known as string like
spikes, with zero energy level and the supersymmetry. This property
seems to prevail under the compactification  of some directions of
the target space, \cite{deWit+Peeters+Plefka}, although a rigorous
proof such as the one given in the case \cite{6} has not yet been
constructed in detail. The spectrum of the CSNW seems to have a
completely different nature according to the results of \cite{5,1}.
These supermembranes require considering a compact sector of the
target space together with a topological condition on the
configuration space.  In this section we describe the main
ingredients of this construction.

When we consider the $D=11$ supermembrane in the light cone
gauge, \cite{deWit+H+N,deWit+M+N}, the potential is given by
\[
    V(X)=\{X^{\tilde{M}},X^{\tilde{N}}\}^2, \qquad \tilde{M},\tilde{N}=1,\ldots,9,
\]
where
\[
   \{X^{\tilde{M}},X^{\tilde{N}}\}=\frac{\epsilon^ab}{\sqrt{W}}\partial_aX^{\tilde{M}}
   \partial_bX^{\tilde{N}},\qquad a,b=1,2.
\]
The scalar density $\sqrt{W}$ is in place as a consequence of the
partial gauge fixing procedure and we take it to be the induced
volume of a minimal immersion introduced below. Here $X^{\tilde{M}}$
are maps from the compact Riemann  surface $\Sigma$ to the target
space. In the present discussion we assume that $\Sigma$ is a torus
and the target space is $\mathbb{M}_7\times S^1\times S^1$, where
$\mathbb{M}_7$ is the Minkowski space-time of dimension 7. More
general target spaces have been consider in \cite{Bell+AR}. We take
$X^r$, $r=1,2$, to be maps from $\Sigma$ to $S^1\times S^1$ and
$X^m$, $m=3,\ldots,9$ maps from $\Sigma$ to $\mathbb{M}_7$.

For the maps $X^r:\Sigma \longrightarrow S^1$ to be well defined,  they
should satisfy the condition
\[
   \oint_{C_i} \ud X^r=m_{ri} \qquad r=1,2,
\]
where $C_i$ is a basis of homology over $\Sigma$ and $m_{ri}$
are integers that depend on the indices $r$ and $i$.  Furthermore,
for the images of $\Sigma$ under $X^r$ to describe a torus we should
also impose the constraint
\begin{equation} \label{e30}
   \int _{\sigma} (\ud X^r \wedge \ud X^s) \epsilon _{rs}=2\pi n\not=0
\end{equation}
 where $n=\det m_{ri}$ for $r=1,2$ and $i=1,2$.  This condition corresponds
 to having a non-trivial central charge on the supersymmetric algebra of the
 supermembrane.

 Among all maps from $\Sigma$ to the target space satisfying the topological
 condition \eqref{e30}, there is one which minimizes the hamiltonian
 of the CSNW. This minimizer realizes in terms of the basis of harmonic one-form
 over $\sigma$, $\ud \widehat{X}^r$. Any one-form over $\sigma$ is given by
 \begin{equation} \label{e31}
    \ud X^r=m^r_s \ud \widehat{X}^s+\delta^{rs} \ud A_s\qquad r,s=1,2
 \end{equation}
 where $A_s$ are single-valued over $\sigma$ and $m^r_s$ are integers. The map
 $X^r$ defined by \eqref{e31} satisfies \eqref{e30}, whenever
 $
     \det m^r_s=1.
 $
 Moreover, using the residual gauge invariance, the area preserving diffeomorphisms
 which are not connected to the identity, we may fix
 $
     m^r_s=\delta^r_s.
 $
 We are then still left with the diffeomorphisms  connected to the identity as a gauge
 invariance of the theory. The transverse coordinates $X^m$ are valued over
 $D=7$ Minkowski space, hence these must be single-valued over $\Sigma$.

Under the above considerations, the  hamiltonian of the CSNW model can now be rewritten in terms of $X^m$ and $A^r$. The resulting expression may be found
in closed form,
\begin{align}
 \tilde{H}=\int_{\Sigma}&(1/2)\sqrt{W}[ (P_{m})^{2}+(\Pi_{r})^{2}+
 (1/2)W\{X^{m},X^{n}\}^{2}
 +W(\mathcal{D}_r X^{m})^{2}+ \nonumber
 \\ & +(1/2)W(\mathcal{F}_{rs})^{2}]+
 \int_{\Sigma}[(1/8)\sqrt{W}n^{2}
 -\Lambda(\mathcal{D}_{r}\Pi_{r}+\{X^{m},P_{m}\})]+
 \label{e32}    \\&
+\int_{\Sigma} \sqrt{W} [- \overline{\psi}\Gamma_{-} \Gamma_{r}
\mathcal{D}_{r}\psi +
 \overline{\psi}\Gamma_{-} \Gamma_{m}\{X^{m},\psi\} +
 \Lambda \{ \overline{\psi}\Gamma_{-},\psi\}] \nonumber
\end{align}
 where $P_m$ and $\Pi_r$ are the conjugate momenta to $X^m$ and
 $A_r$ respectively. Here $\mathcal{D}_r$ and $\mathcal{F}_{rs}$
 are the covariant derivative and curvature, respectively, of a
 symplectic connection \cite{4}
 constructed from the symplectic structure $\frac{\epsilon^{ab}}{\sqrt{W}}$
 introduced by the central charge, where $\sqrt{W}$ is the volume
 induced
 by the minimal immersion $\widehat{X}^r$.

By using the fact that the explicit expression of $\tilde{H}$  is given
completely in terms of single-valued objects over $\Sigma$, one may find the regularization \cite{2}, $H$. Here the non-exact modes that appear in the
regularization of the $D=11$ supermembrane on compact target space
described in \cite{deWit+Peeters+Plefka} are not an issue, since those modes may be gauge fixed as a consequence of the topological condition. The residual gauge
symmetry, the diffeomorphisms connected to the identity, are described in terms
 of the symplectic non-commutative Yang-Mills formulation
 \eqref{e32}. In the explicit expression, the term
$\Lambda(\mathcal{D}_r\Pi_r+\{X^m,P_m\})$ where $\Lambda$ is a Lagrange
 multiplier, describes  the generator (Gauss Law) of the symmetry.

 The regularized
 hamiltonian realizes as a self-adjoint operator acting on $L^2(\R^N,\C^d)$,
 where $N$ and $d$ are large integers. In this realization, the regularized
bosonic potential is given explicitly by
\[
 \tilde{V}_B(X,\CA) = V_1(X) + V_2(\CA) + V_3(X,\CA)
\]
where
\begin{align*}
   V_1(X)&= 4\sum_{D,m,n} \left|\kappa\sum_{B,C} \sin \left(\frac{B\times C}{\kappa} \pi\right)
   X^{Bm}X^{Cn} \delta ^D_{B+C}\right|^2, \\
   V_2(\CA)&=  4\sum_{r,s,D} \left|\omega^{(D\times V_r)/2}
   \kappa \sin \left(\frac{V_r\times D}{\kappa}
   \pi\right)\CA_s^{D}  -\omega^{(D\times V_s)/2}\kappa
   \sin \left(\frac{V_s\times D}{\kappa} \pi\right)
   \CA_r^D \right.\\  &\left. +i
   \sum_{B,C} \kappa \sin \left(\frac{B\times C}{\kappa} \pi\right)
   \CA_r^B \CA_s^C \delta_{B+C}^D \right|^2,\\
   V_3(X,&\CA)= 2 \sum_{D,r,m}\left|\omega^{(D\times V_r)/2}
   \kappa \sin \left(\frac{V_r\times D}{\kappa}
   \pi\right) X^{Dm} \right.\\ & \left.+ i \sum_{B,C}
   \kappa \sin \left(\frac{B\times C}{\kappa} \pi\right)\CA^B_rX^{Cm} \delta
   ^D_{B+C}\right|^2.
   \end{align*}
Here  $\omega=e^{2\pi i/\kappa}$, where $\kappa$ is a large integer and the indices
\begin{gather*}
    B,C,D\in \{(a_1,a_2)\,:\,a_j=0,\ldots,\kappa-1,\, (a_1,a_2)\not=(0,0)\}, \\
    m,n\in \{1,\ldots,7\}, \qquad
    r,s\in \{1,2\}, \\
    V_1=(1,0), \,V_2=(0,1) \quad \mathrm{and} \quad
    (a_1,a_2)\times (b_1,b_2)=a_1b_2-a_2b_1.
\end{gather*}
The non-zero components of $\mathcal{A}_r$ are
$\mathcal{A}_1^{(a1,0)}$ when $a_1\not=0$ and
$\mathcal{A}_2^{(a_1,a_2)}$ when $a_2\not=0$, while the other
components may be gauge fixed to zero \cite{2}.

The  following lemma was originally formulated in
\cite[Lemmas~1~and~2]{5}.

\begin{lemma} \label{t4}
$\tilde{V}_B$ only vanishes at the origin and
there exist constants $\tilde{a},M>0$ such that
\begin{equation} \label{e10}
  \tilde{V}_B(X,\mathcal{A})+\tilde{a} \geq M (|X|^2+|\mathcal{A}|^2),
  \quad  (X,\mathcal{A})\in \R^N.
\end{equation}
\end{lemma}

\proof The gauge fixing conditions ensure the first part of the
lemma, cf. \cite[Lemma~1]{5}. For the second part, let $\mathbb{T}$
be the unit ball of $\R^N$. We write $X^{Bm}$ and $A^B_r$ in polar
coordinates as
\[
 X^{Bm}=R\phi^{Bm}, \, \CA^B_r=R \psi^B_r \qquad \mathrm{or}
 \qquad X=R\phi, \, \CA=R \psi
\]
where $R\geq 0$, $\phi=(\phi^{Bm})$, $\psi=(\psi^B_r)$ and
$(\phi,\psi)\in \mathbb{T}$.  A straightforward calculation yields
\[
  \tilde{V}_B(R\phi,R\psi)=R^4k_1(\phi,\psi)+R^3k_2(\phi,\psi) +
  R^2k_3(\phi,\psi),
\]
where $k_1(\phi,\psi) \geq 0$, $k_3(\phi,\psi) \geq 0$,
$k_2(\phi,\psi)\in \R$ and it is allowed to be negative but if $k_1(\phi,\psi)=0$,
then $k_2(\phi,\psi)=0$. If $k_1(\phi,\psi)=k_2(\phi,\psi)=0$, then necessarily $k_3(\phi,\psi)\not=0$. The $k_j$ are continuous in
$(\phi,\psi)\in \mathbb{T}$.
When clear from the context, we write $k_j\equiv
k_j(\phi,\psi)$.

Let  $P_{\phi,\psi}(R):=R^2k_1+Rk_2+
   k_3$, so that   $\tilde{V}_B(R\phi,R\psi)=R^2P_{\phi,\psi}(R)$.
Then $P_{\phi,\psi}(R)$ is a family of parabolas parameterized by
$(\phi,\psi)\in \mathbb{T}$.
Clearly $P_{\phi,\psi}(R)>0$ for $R>0$, otherwise the first part of the lemma
is violated.  Hence the desired property follows by  noticing that
for $R\geq 1$,
\[
 \min_{(\phi,\psi)\in\mathbb{T}} \tilde{V}_B(R\phi,R\psi)=
 R^2 \min_{(\phi,\psi)\in\mathbb{T}} P_{\phi,\psi}(R) \geq R^2
 \min_{(\phi,\psi)\in\mathbb{T}} \mu(\phi,\psi),
\]
where
$
 \mu(\phi,\psi):= \min_{R\geq 1} P_{\phi,\psi}(R)>0
$
is a continuous function of $\mathbb{T}$.  This completes the proof of the lemma.

\medskip

In order to simplify out subsequent arguments, we translate the
bosonic potential of the CSNW by a factor of $\tilde{a}$ and write
$V_B:=\tilde{V}_B+\tilde{a}$. The conditions on $V$ in the
hypothesis of Theorem~\ref{t1} below, ensure that this is completely
harmless.


\section{Trace ideals and the bosonic heat kernel}

In this section we show that $e^{-H_Bt}$ has finite trace. We first
review some standard mathematical facts regarding the theory of
trace ideals and one-parameter semigroups, and provide rigorous
definition of the operators $H_B$ and $H$.

Let $1\leq r <\infty$. We recall that a compact operator $T$ is
said to be in the $r^{\mathrm{th}}$ von Neumann-Schatten Class
$\CC_r$ if, and only if,
\[
    \|T\|_r := \left( \sum_{n=0}^{\infty} \mu_n^{r/2}
    \right)^{1/r}< \infty
\]
where $\mu_n$ are the singular values of $T$.
The classes $\CC_1$ and $\CC_2$
are the operators of finite trace
and of Hilbert-Schmidt type respectively, cf. \cite{res1}. If $T$ is
an integral
operator acting on $L^2(\R^N;\C^d)$ with kernel
$K(x,y)\in \C^{d\times d}$, then
\begin{equation*}
   \|T\|^2_2=\int _{(x,y)\in \R\times \R}\!\!\!\! \mathrm{Tr}\,
   [K(x,y)
   K(x,y)^\ast ] \ud x \ud y ,
\end{equation*}
where here the asterisk denotes conjugated transpose of the corresponding
matrix. Notice that $\|T\|\leq \|T\|_r$ for all $1\leq r <\infty$.
The sum of two operators in $\CC_r$ also lies
in $\CC_r$. In fact, the normed linear spaces $\left(\CC_r,\|\!
\cdot \!\|_r\right)$ are complete so these are complex Banach
spaces. Furthermore $\CC_r$ are ideals with respect to the product
in the algebra of all bounded linear operators and $\CC_r\subset
\CC_s$ whenever $r<s$. See \cite[ch.XI.9]{ds2} or \cite{res1} for
references to the proofs of these results and a more complete
account on the elementary properties of trace ideals. In
particular we will employ below the following well known facts.
\smallskip \newline
- Duality: if $T\in \CC_r$, $S\in \CC_s$ and $q^{-1}=r^{-1}+s^{-1}$,
where
$q,r,s\geq 1$, then
\begin{equation} \label{e2}
\|TS\|_q\leq \|T\|_r \|S\|_s,
\end{equation}
\newline - Interpolation: if $T\in \CC_2$ and
$r>2$, then
\begin{equation} \label{e1}
   \|T\|_r\leq \|T\|_2^{2/r}\|T\|^{1-2/r}.
\end{equation}
\newline
Notice that \eqref{e2}
ensures that any c$_0$ one-parameter
semigroup lying on $\CC_r$ for some  $r>1$,
should also lie on $\CC_1$.

\medskip

Below and elsewhere we assume that $N$ and $d$
are fixed positive integers, and $M$ is as in  Lemma~\ref{t4}.  In order to simplify the notation in our subsequent analysis, we rearrange the spacial coordinates and denote $(X,\mathcal{A})\equiv x\in \R^N$.

Let $H_A:=(-\Delta+c|x|^2)\otimes \mathbb{I}$ with domain
\[
   \dom H_A=H^2(\R^N;\C^d)\cap \widehat {H^2(\R^N;\C^d)}\subset
   L^2(\R^N;\C^d),
\] where $c>0$ is a small constant. A reasoning
involving the isometries $U_\alpha$
introduced below, easily shows that this operator
is self-adjoint and positive in the above domain. The
eigenvalues of $H_A$ can be computed explicitly and
$\|e^{-H_At}\|< 1$ for all $t>0$.

The heat kernel of $H_A$ is found explicitly as follows.
Let the unitary group of isometries
$U_\alpha \phi(x):=e^{N\alpha/2}\phi(e^\alpha x)$, $\alpha\in \R$.
Since
\[
   U_\alpha [(-\Delta+|x|^2)\otimes \mathbb{I}]
   U_{-\alpha}\phi(x)=[e^{-2\alpha}(-\Delta+e^{4\alpha}|x|^2)
   \otimes \mathbb{I}]\phi(x)
\]
for all $\phi\in \dom H_A$ and
\[
   e^{-U_\alpha (P^2+Q^2)U_{-\alpha}t}\phi(x)=U_\alpha
   e^{-(P^2+Q^2)t}U_{-\alpha}\phi(x),
\]
by virtue of Mehler's formula, \eqref{e9},
\begin{gather*}
 K_t^A(x,y)=(w_1/\pi)^{N/2}\exp[2w_1(x\cdot y)-w_2(|x|^2+|y|^2)]
 \,\mathbb{I},
 \\
 w_1=\frac{c^{1/2}\lambda}{1-\lambda^2},\qquad w_2=c^{1/2}\frac{1+\lambda^2}
 {2(1-\lambda^2)},\qquad \mathrm{and}\qquad
 \lambda=\exp[-2 c^{1/2}t].
\end{gather*}
Moreover $e^{-H_At}$ is Hilbert-Schmidt for all
$t>0$. Indeed,
\begin{align*}
   \|&e^{-H_At}\|_2^2 = \int_{\R^N\times \R^N} |K^A_t(x,y)|^2 \ud x \ud y \\
   & =  (w_1/\pi)^N \int \exp[-2 (w_2^2-w_1^2)|x|^2/w_2]
   \exp[-2w_2|y-(w_1/w_2)x|^2] \ud x \ud y \\
    &= (w_1/\pi)^N \int \exp[-|x|^2/(2w_2)]\ud x \int
   \exp[-2w_2|y|^2] \ud y \\
   & = w_1^{N}<\infty.
\end{align*}
In particular, $e^{-H_At}$ is a compact operator so by virtue of
the spectral mapping theorem, we can compute explicitly
$\|e^{-H_At}\|=e^{-Nc^{1/2}t}$. Notice that $Nc^{1/2}$ is the
ground eigenvalue of $H_A$.

\medskip

Let us now consider
$H_B:=(-\Delta+V_B)\otimes \mathbb{I}$. By virtue of lemma~\ref{t4},
$\dom H_B\subseteq \dom H_A$, then we may define rigorously the
domain of $H_B$ by means
of Friedrichs extensions techniques, cf. \cite{res2}. In
this domain $H_B$ is self-adjoint.
Furthermore, variational arguments along with
\eqref{e10}, ensure that $H_B\geq 0$ and it has a
complete set of
eigenfunctions whose eigenvalues accumulate at $+\infty$.

By virtue of the Feynman-Kac formula (cf. \cite{glja}),
by putting $c= M$ in the expression for $H_A$, we achieve
\begin{equation} \label{e6}
0<K_t^B(x,y)\leq K^A_t(x,y)\ \mathrm{for\ all}\ x,y\in
\R^N\ \mathrm{and}\ t>0.
\end{equation}
Thus
\begin{equation} \label{e12}
   \|e^{-H_Bt}\|_2^2=e^{-bt}\int_{\R^N\times \R^N }|K^B_t(x,y)|^2 \ud x \ud y\leq
   w_1^N,
\end{equation}
so that $e^{-H_Bt}$ is also a Hilbert-Schmidt operator and hence
it has finite trace. Moreover, since $H_B\geq H_A$, the
$k^\mathrm{th}$ eigenvalue
of $H_B$ is bounded below by the $k^\mathrm{th}$ eigenvalue
of $H_A$ and $\|e^{-H_Bt}\| \leq e^{-NM^{1/2}t}<1$ for all $t>0$.

\medskip

We conclude this section by examining the full hamiltonian $H$.
According to \cite[\S3-5]{1}, $H$ is a relatively bounded
perturbation of $H_B$,
$H:=H_B+V_F$, where  $V_F(x)$ is a Hermitian
$d\times d$ matrix-valued function of $\R^N$ linear in the variable $x$.
Thus the $(i,j)-$th entry of $V_F$ satisfies
\begin{equation} \label{e8}
   |V_F^{ij}(x)|\leq a(1+|x|),\qquad \qquad x\in \R^N,
\end{equation}
for some constant $a$ independent of $i,j$. This identity will play a crucial role
in the description of the heat kernel of $H$.
In physical terms, the potential $V_F$ comprises  the fermionic
contribution in the CSNW model which, by the very special
characteristics of
this representation, happens to be dominated by the bosonic section
of the theory.

By virtue of \eqref{e10} and \eqref{e8},
the operator of multiplication by $V_F$ is relatively
bounded with respect to $H_B$ with relative bound
zero. Hence, the standard argument,
see e.g.  \cite[Theorem~1.1, p.190]{kat}, shows that
$H$ is a self-adjoint
operator if we choose $\dom H:=\dom H_B$.

In \cite{1} a variational procedure is employed in order to
show that $H$ is bounded below, it possess a complete set of
eigenfunctions and its spectrum comprises a discrete set of
eigenvalues whose only accumulation point
is $+\infty$.
Notice that the spectral theorem together
with Theorem~\ref{t1} below, yields to an independent proof of this fact.


\section{The Dyson expansion for $e^{-Ht}$}

By employing a well known result due to Hille and Phillips,
see \cite[theorem 13.4.1]{hipi}, in this
section we construct a series expansion for $e^{-Ht}$ in terms
of $e^{-H_Bt}$ and $V_F$. We then prove that
Lemma~\ref{t4} and \eqref{e8} ensure convergence of this series for
all $t>0$ in the topology of the von Neumann-Schatten class.
Notice that Theorem~\ref{t1}, the main result of this section,
may easily be formulated
for general one-parameter semigroups arising from
matrix integral operators. Nonetheless, in order to avoid distractions from our main
purpose, we choose to restrict our notation to the concrete
situation under discussion.

The following criterion on perturbation of generators
of one-parameter semigroups may be found in
\cite[theorem 13.4.1]{hipi}, see also \cite[\S3.1]{ops}.
Let $V:\R^N\longrightarrow \C^{d\times d}$ be a potential
such that
\begin{equation}\label{e3}
  \int_0^1 \|Ve^{-H_Bt}\| \ud t<\infty
\end{equation}
where, by hypothesis, $e^{-H_Bt} \phi$ lies in the domain
of closure of $V$ for all $\phi\in L^2(\R^N;\C^d)$ and $t>0$. Then
$H_B+V$ is the
generator of a one-parameter semigroup $e^{-(H_B+V)t}$ for $t>0$.
Furthermore, the operator-valued functions
\begin{equation} \label{e4}
\begin{aligned}
 W_0(t) \phi &:= e^{-H_Bt}\phi, \\
 W_1(t) \phi & := -\int _{s_1=0} ^t  e^{-H_B(t-s_1)} V e^{-H_Bs_1}
 \phi \ud s_1, \\
 W_2(t) \phi & := \int _{s_1=0} ^t \int _{s_2=0}^{s_1} e^{-H_B(t-s_1)}
 Ve^{-H_B(s_1-s_2)} V e^{-H_Bs_2} \phi \ud s_2 \ud s_1, \\&
 \vdots \\
 W_k(t)\phi & := (-1)^k\int _{s_1=0}
 ^t \cdots \int _{s_{k}=0}^{s_{k-1}}
 e^{-H_B(t-s_1)} Ve^{-H_B(s_1-s_2)} V \ldots \\
 &\hspace{.5in} \ldots e^{-H_B(s_{k-1}-s_k)} V e^{-H_Bs_k} \phi \ud s_k
 \ldots \ud s_1,
\end{aligned}
\end{equation}
 are well defined and there is a $\delta>0$, small enough, such that
\begin{equation} \label{e7}
   e^{-(H_B+V)t}\phi = \sum_{k=0}^\infty W_k(t)\phi
\end{equation}
for all $0<t<\delta$. Under
hypothesis \eqref{e3}, the
convergence of both the integrals and
the series above is only guaranteed in the strong operator
topology. This is Dyson's series for the heat kernel of a
perturbed hamiltonian.

In order to recover the heat kernel of $H$, we may proceed as
follows. Since
$e^{-H_Bt}$ is given by the kernel $K^B_t(x,y)$, a
straightforward computation yields
\[
   W_k(t)\phi(x)=\int_{y\in \R^N} K_{t,k}(x,y) \phi(y) \ud y,
\]
where $K_{t,0}(x,y)=K^B_t(x,y)$ and
\begin{align*}
   K_{t,k}(x,y)&:= -\int_{0}^t \int _{\R^N} K^B_{t-s}(x,z)
   V(z) K_{s,k-1}(z,y) \ud z\! \ud s.
\end{align*}
We remark that here $K_{t,k}(x,y)$ is a $d\times d$ matrix.
The convergence
in the strong operator topology ensures that
\[
   \lim_{n\to \infty}\int_{\R^N}
   \Big[K_t(x,y)-
   \sum_{k=0}^n K_{t,k}(x,y)\Big] \phi(y) \ud y=0, \qquad t<\delta,
\]
for all $\phi\in L^2(\R^N;\C^d)$.

\medskip

Although the above result may be useful in some applications,
\eqref{e3} only guarantees convergence in the strong operator
topology for small $t>0$. This provides a far from satisfactory
description of $K_t(x,y)$.
The following theorem shows that under more restrictive hypotheses
on $V$, including those satisfied by $V_F$,
uniform convergence is guaranteed for all $t>0$ is the
much stronger  topology of the spaces $\CC_r$.

\begin{theorem} \label{t1}
Let $V:\R^N\longrightarrow
\C^{d\times d}$ be a potential such that
\begin{equation} \label{e11}
   \max_{i,j=1,\ldots, d} |V_{ij}(x)|\leq a(1+|x|^\alpha)\qquad
   \qquad x\in \R^N
\end{equation}
for constants $a>0$ and $0\leq \alpha <2$. Let $r>2N/(2-\alpha)$.
Then $W_k(t)\in \CC_r$ and
$\sum_{k=0}^\infty \| W_k(t)\|_r<\infty$ for all $t>0$. Hence
\begin{equation*}
e^{-(H_B+V)t}=\sum_{k=0}^\infty W_k(t)\in \CC_r,
\end{equation*}
where the series converges uniformly in the norm $\|\!\cdot\!\|_r$
for all $t>0$.
\end{theorem}

\proof
\underline{Step 1}: we first show that
\begin{equation} \label{e5}
\|Ve^{-H_Bt}\|_r\leq b t^{-1+\varepsilon} \qquad \mathrm{for\ all}
\quad t>0,
\end{equation}
where $b>0$ and $0<\varepsilon<1$ are constants
independent of $t$.
Below and elsewhere $a_j$ are constants
independent of $t$ or $n$,
but might depend on other parameters such as
$r$, $d$, $p$ or $N$.

For $n\in \N$, let
\[
   \Lambda_n=\{x\in \R^N: 2^n<|x|<2^{n+1}\}
\]
and denote by $\chi_n(x)$ the characteristic function of this set.
Let $p>1$ be a fixed parameter. By virtue of \eqref{e6},
\begin{align*}
   \|\chi_n e^{-H_Bt}\|_2^2 &=\int_{\R^N\times \R^N} |\chi_n(x)|^2|K_t^B(x,y)|^2 \ud
   x \ud y \\
   &\leq a_1 w_1^N\int _{\Lambda_n} \exp [-|x|^2/(2w_2)] \ud x
   \int _{\R^N} \exp [-2w_2|y|^2] \ud y\\
   & = a_2 \left(\frac{w_1}{w_2^{1/2}}\right)^N
  \int _{\Lambda_n} \exp [-|x|^2/(2w_2)] \ud x \\
  &=a_3 \left(\frac{w_1}{w_2^{1/2}}\right)^N \int_{2^n}^{2^{n+1}}
  r^{N-1}\exp [-r^2 /(2w_2)]\ud r \\
  &\leq a_4 \left(\frac{w_1}{w_2^{1/2}}\right)^N \int_{2^n}^{2^{n+1}}
  (r/(2w_2)^{1/2})^{-(p+1)} r^{N-1} \ud r \\
  & \leq a_5 w_1^N w_2^{(p+1-N)/2}   2^{-n(p+1-N)}.
\end{align*}
Notice that both $w_1$ and $w_2$ are continuous in $t$.
Furthermore
\begin{gather*} w_1\sim t^{-1},\qquad w_2\sim t^{-1}\qquad
\mathrm{as}\qquad t\to 0, \ \mathrm{and} \\
w_1\sim e^{-2M^{1/2}t}, \qquad w_2\sim 1\qquad
\mathrm{as}\qquad
t\to \infty.
\end{gather*}
Also $\|\chi_n e^{-H_Bt}\| \leq \|e^{-H_Bt}\| \leq 1$
for all $t>0$. Then, \eqref{e1} yields
\begin{equation*}
  \|\chi_n e^{-H_Bt}\|_r \leq a_6 t^{-(p+1+N)/(2r)} 2^{-n(p+1-N)/r}
\end{equation*}
for all $t>0$.

Let $\tilde{V}(x):=\sum_{n=0}^\infty
2^{\alpha(n+1)} \chi_n(x)$. Then
\[
  \|\tilde{V}e^{-H_Bt}\|_r\leq a_7 t^{-(p+1+N)/(2r)}
  \sum_{n=0}^\infty 2^{n[\alpha-(p+1-N)/r]}.
\]
Thus, if $r$ and $p$ are large enough, such that
\[
   0<\frac{p+1+N}{2r}<1 \qquad \mathrm{and} \qquad
   \frac{p+1-N}{r}>\alpha,
\]
estimate \eqref{e5} holds for $\tilde{V}$.
This is possible only when $0\leq \alpha <2$,
$
    r>2N/(2-\alpha)$ and
    $ p> \frac{\alpha+2}{2-\alpha}N-1$,
the first two conditions being precisely to the ones required
in the hypothesis the theorem. By virtue of \eqref{e1} and \eqref{e12},
\[
  \|e^{-H_Bt}\|_r\leq w_1^{N/r}\|e^{-H_Bt}\|^{1-2/r}
  \leq a_{8}t^{(\alpha-2)/\alpha}, \qquad t>0.
\]
Then we may achieve \eqref{e5} for $V$, by observing that
\begin{equation*} \label{e16}
\begin{aligned}
  \|Ve^{-H_Bt}\|_r&=\|V(\tilde{V}+1)^{-1}(\tilde{V}+1)e^{-H_Bt}\|_r
  \\ &\leq \|V(\tilde{V}+1)^{-1}\|
  \|(\tilde{V}+1)e^{-H_Bt}\|_r \\ & \leq
   a_{9} \sup_{x\in \R^N}
   \max_{i,j=1,\ldots,d} \frac{|V_{i,j}(x)|}{\tilde{V}(x)+1}
    (\|\tilde{V}e^{-H_Bt}\|_r+\|e^{-H_Bt}\|_r) \\
    &\leq a_{10} (\|\tilde{V}e^{-H_Bt}\|_r+\|e^{-H_Bt}\|_r)
\end{aligned}
\end{equation*}
for all $t>0$.

From \eqref{e5} it follows directly that \eqref{e3} holds.
Then the operators $W_k$ in \eqref{e4} are well defined
and $e^{-(H_B+V)t}$ is
given by expression \eqref{e7} for $0<t<\delta$.

\smallskip

\underline{Step 2}: we show that $e^{-H_Bt}$ is continuous
for $t>0$ in the
norm $\|\cdot\|_r$. Let
$\{\nu_1\leq\nu_2\leq \ldots\}\subset (0,\infty)$
be the spectrum of $H_B$. Since $e^{-H_Bt}$ is a compact operator,
the spectrum of $e^{-H_Bt}$
comprises the points $\{0\}\cup\{\ldots\leq e^{-\nu_2 t}\leq  e^{-\nu_1
t}\}$ and
\[
   \|e^{-H_Bt}\|_r^r = \sum_{k=1}^\infty e^{-r\nu_kt}<\infty.
\]
Since $e^{-H_Bt}$ and $e^{-H_Bs}$
have the same eigenfunctions, by virtue of the spectral mapping theorem
and the dominated convergence theorem,
\[
  \|e^{-H_Bt}-e^{-H_Bs}\|_r^r=\sum_{k=1}^\infty
  |e^{-\nu_kt}-e^{-\nu_ks}|^r\to 0 \qquad \mathrm{as} \quad
|s-t|\to 0
\]
so $e^{-H_Bt}$ is continuous in $\|\!\cdot\!\|_r$.

\smallskip

\underline{Step 3}: let us show that the conclusion of the
theorem holds. In Section~2 we saw that the unperturbed semigroup
$W_0(t)=e^{-H_Bt}\in \CC_1\subset \CC_r$ for all $t>0$.
In order to achieve the same conclusion for the remaining $W_k$
we proceed as follows.

For $k=1$, Step~2 ensures that
 $e^{-H_B(t-s_1)} V e^{-H_Bs_1}$ is
continuous in $\CC_r$ for $s_1>0$. Since
\begin{align*}
 \int_{s_1=0}^t \|e^{-H_B(t-s_1)} &V e^{-H_Bs_1}\|_r \ud s_1 \leq
     \int_{s_1=0}^t \|V e^{-H_Bs_1}\|_r \ud s_1 \\
    &<\frac{bt^\varepsilon}
    {\varepsilon} =:b(1) < \infty,
\end{align*}
it is also integrable in
$\CC_r$.
Thus the (Riemann) integral
in the definition of $W_1(t)$ converges in $\|\!\cdot\!\|_r$ and so
$W_1(t) \in \CC_r$ for all $t>0$.

Let $c(k)=\int_0^1(1-u)^{-1+\varepsilon}u^{-1+k\varepsilon}\ud u$.
For $k=2$, notice that the $\CC_r$-valued function $e^{-H_B(t-s)}
Ve^{-H_B(s-u)} V e^{-H_Bu}$ is $\|\cdot\|_r$-continuous in both
$s$ and $u$ for all $0<u<s<t$. Furthermore,
\begin{align*}
 \int _{s_1=0} ^t \int _{s_2=0}^{s_1} \| e^{-H_B(t-s_1)} &
 Ve^{-H_B(s_1-s_2)} V
 e^{-H_Bs_2}\|_r
  \ud s_2 \ud s_1  \leq \\
  & \leq  \int _{s_1=0} ^t \int _{s_2=0}^{s_1} \| V e^{-H_B(s_1-s_2)}\|_r
\|V e^{-H_Bs_2}\|_r
  \ud s_2
  \ud s_1 \\ 
  & \leq  b ^2 \int _{s_1=0} ^t \int _{s_2=0}^{s_1}
   (s_1-s_2)^{-1+\varepsilon} s_2^{-1+\varepsilon}\ud s_2 \ud s_1 \\
    & =  b^2\int _{s_1=0} ^ts_1^{-1+2\varepsilon}\ud s_1
  \int_{u=0}^1 (1-u)^{-1+\varepsilon} u ^{-1+\varepsilon}\ud u  \\
  & =  \frac{b^2t^{2\varepsilon}}{2\varepsilon}c(1)<\infty .
\end{align*}
Then, the integral in the definition of $W_2(t)$
also converges in the $\CC_r$-norm and so $W_2(t)\in \CC_r$ where
\[
   \|W_2(t)\|_r \leq  \frac{b^2t^{2\varepsilon}}{2\varepsilon}c(1)
\]
for all $t>0$.

Similar computations show that $W_k(t)\in \CC_r$  and
\[
   \|W_k(t)\|_r \leq \frac{b^kt^{k\varepsilon}}{k\varepsilon}
   c(1)c(2)\cdots c(k-1)=:b(k)
\]
for all $k=3,4,\ldots$ and $t>0$.
Since $\sum_1^\infty b(k)<\infty$,  the right hand side of
\eqref{e6} converges in the $\CC_q$-norm for all $t>0$.
This completes the proof of Theorem~\ref{t1}.

\medskip

In particular the conclusions of the above theorem hold
when $V\equiv V_F$ so that $H$ is the regularized hamiltonian
of the CSNW model. Let $\{\lambda_1\leq \lambda_2\leq \ldots \}$
be the set of eigenvalues
of $H$. Since $e^{-Ht}\in \CC_1$, $\sum_{k=1}^\infty e^{-\lambda_nt}
<\infty$ for all $t>0$.
This remarkable result provides information about the asymptotic behaviour
of $\lambda_n$ as $n\to \infty$.

\section{A Feynman-Kac description of the heat kernel}

In this final section we discuss various aspects of the
Feynman-Kac integral formulation of $K_t(x,y)$.

\subsection{A formulation \textit{via} standard Wiener measures}
Let us first examine the standard line of reasoning
 that leads toward the formulation
of a Feynman-Kac
identity for the heat kernel of $H$, cf.
\cite{glja}. For this we decompose
$H=P^2+V_S$, where $V_S(x)=V_B(x)\otimes \mathbb{I}+V_F(x)$, i.e.
we regard the regularized hamiltonian as a perturbation of the
free hamiltonian $P^2\equiv -\Delta\otimes \mathbb{I}$.
According to the results of \cite{1}, the eigenvalues of the
overall potential matrix $V_S(x)$ are smooth functions
of the variable  of the configuration space, $x$, and they
diverge to infinity as $x\to \infty$.
This ensures that the potential operator is bounded
from below. The Trotter-Kato
formula, cf. \cite{ops} or \cite{glja}, guarantees that
\begin{equation} \label{e13}
    e^{-Ht}\phi(x)=\lim_{n\to \infty} \left(e^{-(\Delta\otimes \mathbb{I})t/n}
    e^{-V_St/n}\right)^n\phi(x),
\end{equation}
for all $\phi\in L^2(\R^N;\C^d)$. For finite $n\in \N$,
\begin{equation} \label{e14}
   \mathrm{Ker}\left[(e^{-(\Delta\otimes\mathbb{I})t/n}
e^{-V_St/n})^n\right]=\int \prod_{j=1}^n e^{-(jt/n)V_S(x(jt/n))}
    \ud W^t_{x,y},
\end{equation}
cf. \cite[Corollary~3.1.2]{glja},
where the integrand is a matrix with bounded components so it is
integrable. For each continuous path $x(\cdot)$ joining
$x$ and $y$,
\[
  \lim_{n\to \infty} \prod_{j=1}^n e^{-(jt/n)V_S(x(jt/n))}
  =:\exp_{\mathrm{ord}} \left[-\int_0^tV_S(x(s)) \ud s \right]
\]
where, by definition, $\exp_\mathrm{ord}$ denotes
the limit in the left hand side.
The convergence of the finite difference scheme described in
\cite[Sections~3.1-3.5]{rm} guarantees the existence of this limit.

By virtue of the dominated convergence theorem for the Borel measure
$W^t_{x,y}$,
\[
   \lim_{n\to \infty} \int \prod_{j=1}^n e^{-(jt/n)V_S(x(jt/n))}
   \ud W^t_{x,y}
  =\int
  \exp_\mathrm{ord} \left[-\int_0^tV_S(x(s)) \ud s \right]
  \ud W^t_{x,y}.
\]
Thus, identity \eqref{e13} and the dominated convergence
theorem, this time applied to the Lebesgue measure in $\R^N$, ensure
that the left side of \eqref{e14} approaches to $K_t(x,y)$
as $n\to \infty$. Thus
we achieve the Feynman-Kac type identity
\begin{equation} \label{e15}
    K_t(x,y)= \int
  \exp_\mathrm{ord} \left[-\int_0^tV_S(x(s)) \ud s \right]
  \ud W^t_{x,y}.
\end{equation}
Since all of the matrix functions involved are continuous,
the above limits are guaranteed to exist for all $x$ and $y$ in
$\R^n$.

\medskip

\subsection{A formulation \textit{via} non-standard Wiener measures}
We conclude this section by considering an alternative description
of $K_t(x,y)$ based on the results of Section~4. Let $V$ be a
matrix-valued potential satisfying \eqref{e11} and such that
$H=H_B+V$ is bounded below. Without loss of generality we will
assume that $H\geq cI$, for some $c>0$, otherwise we may add a
suitable constant to $H$ and modify conveniently the resulting
identity. Let
\[
   F(t):=e^{-(t/2)H_B}(\mathbb{I}-tV)e^{-(t/2)H_B}.
\]
By virtue of \eqref{e5}, $F(t)$ is a bounded operator
for all $t>0$. It is also self-adjoint, $F(0)=I$
and $F(t)\phi\to \phi$ as $t\to 0$ for all $\phi$. Notice that
$F(t)$ is also a compact operator and, in fact,
it lies in $\CC_r$ for $r>2N/(2-\alpha)$.

A direct calculation yields
\begin{equation} \label{e18}
   \lim_{t\to 0} (t)^{-1}[F(t)\phi-\phi]=-H\phi
\end{equation}
for all $\phi \in \mathrm{Dom}\,H$.

\begin{lemma} \label{t2}
There exists $t_0>0$ independent of $\phi$,
such that $0<(\phi,F(t)\phi)<1$ for all $t>t_0$ and
$\phi\in L^2(\R^N;\C^d)$ with $\|\phi\|=1$.
\end{lemma}
\proof
Let $\phi$ be as in the hypothesis. By virtue of \eqref{e5},
\[
\|e^{-H_Bt/2}Ve^{-H_Bt/2}\|\leq \|e^{-H_Bt/2}Ve^{-H_Bt/2}\|_r
\leq bt^{-1+\varepsilon} \qquad \mathrm{for\ all}
\quad t>0,
\]
where $b$ is independent of $t$. Thus
\[
(\phi,F(t)\phi) \geq \|e^{-H_Bt}\phi\|^2+t^\varepsilon
\leq (1+t^\varepsilon).
\]
On the other hand, since $H\geq cI$,
\[
   (\phi,F(t)\phi) \leq (1-tc)\|e^{-H_Bt}\phi\|^2
   \leq (1-tc).
\]
Hence the lemma is proven.

\medskip

Since $F(t)$ is a self-adjoint operator, $\|F(t)\|<1$ for
all $t>t_0$. It is well known, cf. e.g. \cite[Lemma~3.28]{ops},
that \eqref{e18} together with this condition ensure
the validity of the following.

\begin{corollary} \label{t3} Let $t>0$ be fixed.
For all $\phi\in L^2(\R^N;\C^d)$,
\[
     e^{-Ht}\phi(x)=\lim_{n\to \infty}F(t/n)^n\phi(x).
\]
\end{corollary}

\medskip

We can regard the latter corollary formally as a
Feynman-Kac identity. Indeed, consider the measure
$\tilde{W}^t_{x,y}$ in the space of continuous path constructed in
the same fashion as the Wiener measure but using $K^B_t(x,y)$,
instead of the kernel of the free evolution semigroup $e^{-P^2
t}$, for defining the measure of cylinder sets, cf.
\cite[p.49]{glja}. Then
\[
  \mathrm{Ker}\left[ F(t/n)^n\right](x,y)
  =\int \prod_{j=1}^n [\mathbb{I}-(t/n)V(x(jt/n))]
  \ud \tilde{W}^t_{x,y}.
\]
In the limit $n\to \infty$, the integrand at the right hand side
converges to
$\exp_{\mathrm{ord}}\big[-\int_0^t V(x(s))\ud s\big]$. Hence
\[
   K_t(x,y)= \int \exp_{\mathrm{ord}}\Big[-\int_0^t V(x(s))\ud s
   \Big]
   \ud \tilde{W}^t_{x,y}.
\]

Whenever $V\equiv V_F$,
the latter, with the usual additional factor arising from the
Fadeev-Popov
procedure, is the path integral description of
the regularized hamiltonian of the CSNW model.
Notice that, although $V_F$ is not bounded from below, it is dominated
by the measure $\tilde{W}^t_{x,y}$.


\section*{Conclusion}
We studied the quantization of the regularized hamiltonian of the
compactified $D=11$ supermembrane with non-zero central charge
arising from a non-trivial winding. By showing that $H$ is a
relatively small perturbation of the bosonic hamiltonian, in
Theorem~\ref{t1} of Section~4 we provided a rigorous Dyson expansion
for the heat kernel of this regularized hamiltonian. We demonstrated
the convergence of this series in the topology of the von
Neumann-Schatten class so that  $e^{-Ht}$ is of finite trace. These
results are relevant in the analysis of the heat kernel of the
$SU(N)$ regularized hamiltonian of the CSNW in the limit $N\to
\infty$. The hypothesis of Theorem~\ref{t1} suggests that the
correct ideal to look at is $\CC_\infty$, as $N$ approaches
$\infty$. In Section~5 we discussed the validity of the Feynman path
integral description of the heat kernel. As a consequence of
Corollary~4, we obtained a matrix Feynman-Kac formula.

The results established in this paper may be formulated in a
completely abstract framework, so they may apply to other
supersymmetric models. Nonetheless, our analysis does not include
the important case of the supermembrane immersed on a $D=11$
Minkowski space. This latter supermembrane has a positive
hamiltonian but its potential is not bounded from below. It is
rather unfortunate that, since this hamiltonian  is not a
perturbation of the bosonic contribution, it is still unclear
whether a Feynman path integral formula exists for this case.


\vspace{1in}

\begin{minipage}{2.5in}
 {\scshape $^1$Lyonell Boulton}\\
{\footnotesize Department of Mathematics \& Statistics, \\
University of Calgary, \\
Calgary, AB, \\ Canada T2N 1N4 \\
email: \texttt{lboulton@math.ucalgary.ca}}
\end{minipage}
\qquad
\begin{minipage}{2.3in}
{\scshape $^2$Alvaro Restuccia}\\
{\footnotesize Departmento de F\'\i sica, \\
Universidad Sim\'on Bol\'\i var, \\
Apartado 89000,\\
Caracas 1080-A, Venezuela. \\
email: \texttt{arestu@usb.ve}}
\end{minipage}


\begin{thebibliography}{99}
\bibitem{duff}{\scshape M.~J.~Duff, T.~Inami, C.~N.~Pope,
E.~Sezgin, K.~Stelle,} ``Semiclassical quantization of the
supermembrane'', {\it Nucl.Phys.} 297 (1989) 515.
\bibitem{od}{\scshape A.~A.~Bytsenko, S.~D.~Odintsov},
``Modular invariance in membrane theory'', Phys. Lett. B 245 (1990)
21-25.
\bibitem{1}{\scshape L. Boulton, M.P. Garc\'{\i}a del Moral,
A.~Restuccia}, ``Discreteness of the spectrum of the compactified
$D=11$ supermembrane with non-trivial winding'', {\it Nucl. Phys. B}
 671 (2003) 343-358. \texttt{hep-th/0211047}.
\bibitem{5}{\scshape L. Boulton, M.P.
Garc\'{\i}a del Moral, I.~Mart\'{\i}n, A.~Restuccia}, ``On the
spectrum of a matrix model for the $D=11$ supermembrane compactified
on a torus with non-trivial winding'', {\it Class. Quantum. Grav.}
19(2002) 2951-2959. \texttt{hep-th/0109153}.
\bibitem{2}{\scshape M.P. Garc\'{\i}a del Moral, A. Restuccia},
``On the spectrum of a noncommutative formulation of the D=11
supermembrane with winding'', {\it Phys. Rev. D} 66 (2002) 045023.
\texttt{hep-th/0103261}.
\bibitem{B1} {\scshape D.~Smith}, ``Intersecting brane solutions
in string and M-theory'', \texttt{hep-th/0210157}.
\bibitem{B2} {\scshape J.F.~Cascales, A.M.~Vuranga}, ``Branes
on generalized Calibrated submanifolds'', \texttt{hep-th/0407132}.
\bibitem{B3} {\scshape J.~Gutowski, G.~Papadopoulos, P.K.~Townsend},
 ``Supersymmetry and generalized calibrations'', \textit{Phys. Rev. D}
60 (1999).
\bibitem{A1} {\scshape K.S.~Stelle}, ``Lectures on supergravity
p-branes'', \texttt{hep-th/9701088}.
\bibitem{A2} {\scshape K.S.~Stelle}, ``BPS branes in supergravity'',
\texttt{hep-th/9803116}.
\bibitem{A3} {\scshape J.P.~Gauntlett}, ``Intersecting branes'',
\texttt{hep-th/9705011}.
\bibitem{A4}{\scshape M.J.~Duff, K.S.~Stelle}, ``Multi-membrane
solutions of $D=11$ supergravity'', \textit{Phys. Lett. B} 253
(1991) 113.
\bibitem{S+T}{\scshape E.~Bergshoeff, E.~Sezgin, P.K.~Townsend},
``Supermembranes and eleven-dimensional supergravity'', {\it Phys.
Lett. B} 189 (1987) 75-78.
\bibitem{A6} {\scshape M.~Cvetic, C.M.~Hull}, ``Wrapped branes
and supersymmetry'', \textit{Nucl. Phys. B} 519 (1998) 141.
\bibitem{glja}{\scshape J.~Glim, A.~Jaffe},
{\em Quantum Physics: A Functional Integral Point of View},
Springer, New York, 1996.
\bibitem{6} {\scshape B. de Wit, M. L\"uscher, H. Nicolai}, ``The supermembrane
is unstable'', {\it Nucl. Phys. B} 320 (1989) 135-159.
\bibitem{deWit+H+N} {\scshape B. de Wit, J.~Hoppe, H. Nicolai},
``On the quantum mechanics of supermembranes'', {\it Nucl. Phys. B}
305 (1988) 545-581.
\bibitem{deWit+M+N}{\scshape B. de Wit, U. Marquard, H. Nicolai}, ``Area-preserving
diffeomorphisms and supermembrane Lorentz invariance'', {\it Comm.
Math. Phys.} 128 (1990) 39-62.
\bibitem{3}{\scshape I. Mart\'{\i}n, J. Ovalle, A. Restuccia},
``Stable solution of the double compactified $D=11$ supermembrane
dual'', {\it Phys. Lett. B} 472 (2000) 77-82.
\texttt{hep-th/9909051}.
\bibitem{4}{\scshape I. Mart\'{\i}n, J. Ovalle, A. Restuccia},
``Compactified D=11 supermembranes and symplectic noncommutative
gauge theories'', {\it Phys. Rev. D} 64 (2001) 046001.
\texttt{hep-th/0101236}.
\bibitem{deWit+Peeters+Plefka}{\scshape B. de Wit, K. Peeters,
J.C. Plefka,} ``The supermembrane with winding'', {\it Nucl. Phys.
Proc. Suppl.} 62 (1998) 405-411. \texttt{hep-th/9707261}.
\bibitem{Bell+AR} {\scshape J.~Bellorin, A.~Restuccia},
``Minimal inmersions and the spectrum of supermembranes'',
\texttt{hep-th/0405216}.
\bibitem{res1}{\scshape M. Reed, B. Simon}, {\em Methods of Modern
Mathematical Physics, Volume 1}, Academic Press, New York, 1972.
\bibitem{ds2}{\scshape N. Dunford, J. Schwartz},
{\em Linear Operators, Part II: Spectral Theory}, Interscience, New
York, 1963.
\bibitem{res2}{\scshape M. Reed, B. Simon}, {\em Methods of Modern
Mathematical Physics, Volume 2}, Academic Press, New York, 1975.
\bibitem{kat}{\scshape T.~Kato}, \emph{Perturbation
Theory for Linear Operators}, Springer-Verlag, 1976.
\bibitem{hipi}{\scshape E.~Hille, R.~Phillips}, {\em Functional
Analysis and Semigroups}, American Mathematical Society, Providence,
1957.
\bibitem{ops}{\scshape E. B. Davies}, {\em One-parameter
Semigroups}, Academic Press, London, 1980.
\bibitem{rm}{\scshape R.~Richtmyer, K.~Morton}, {\em
Difference Methods for
Initial-Value Problems}, Interscience, New York, 1967.

\end{thebibliography}
\end{document}